\documentstyle[12pt]{ioplppt} 
 
\date{} 
 
\begin{document} 
\jl{1} 
\title{Symmetry analysis of the 1+1 dimensional relativistic imperfect fluid 
dynamics}[Relativistic imperfect fluid] 
 
\author{C Alexa\dag \,\, and \,\, D Vrinceanu\ddag } 
 
\address{\dag\ Department of Particle Physics, IFIN-HH, Bucharest 76900, RO} 
\address{\ddag\ Dept. of Theoretical Phys. \& Mathematics,  
University of Bucharest, RO} 
 
\begin{abstract}
The flow of the relativistic imperfect fluid in two dimensions is discussed. 
We calculate the symmetry group of the energy-momentum tensor conservation 
equation in the ultrarelativistic limit. Group-invariant solutions for the 
incompressible fluid are obtained. 
\end{abstract} 
 
\maketitle 
 
\section{Introduction} 
 
Many physical systems may be approximately regarded as perfect fluids. A 
perfect fluid is defined \cite{wei} as having at each point a velocity  
${\it \vec v}$, such that an observer moving with this velocity sees the fluid 
around him as isotropic. This will be the case if the mean free path between 
collisions is small compared with the scale of lengths used by the observer. 
But one often has to deal with somewhat imperfect fluids, in which the 
pressure or velocity vary appreciable over distances of the order of a mean 
free path, or over times of the order of a mean free time, or both. 
 
Numerical methods to solve the hydrodynamical equations have been  
discussed for instance in \cite{rischke}. 
Any attempt to solve numerically the relativistic hydrodynamics equations is  
hardly discouraged by the puzzle of choosing the rest frame; this situation  
leads us to acausal and instable solutions \cite{str}.  
 
It is therefore of interest to use different methods that are directly 
related to the solutions of the equations of the relativistic fluid 
dynamics. Important information may be achieved using the Lie symmetry group 
of the covariant relativistic hydrodynamics equations. Symmetry analysis is 
one of the systematic and accurate ways to obtain solutions of differential 
equations. The power of this technique consists in the possibility to 
explore the properties of physical systems like the symmetry structure and 
the invariants and then solving the corresponding reduced differential  
equations. 
Interesting systems were successfully studied using this approach for 
example \cite{olver,ludu}. In \cite{ale} we have already investigate a 
particular simple form of the energy-momentum tensor conservation equation. 
 
In the next section we briefly discuss the general relativistic fluid 
formalism, the ultrarelativistic approximation and the final form of the 
energy-momentum tensor conservation equations. In Sec. 3 we address the 
symmetry group of transformations and its Lie algebra. Sec. 4 is devoted to 
integrability conditions, invariants and Sec. 5 to group invariant-solutions 
analysis. In the last section we present miscellaneous comments and final 
conclusions. 
 
\section{Energy-momentum tensor} 
 
Relativistic fluid dynamics is well describe by the number of particles N 
and energy-momentum tensor $T_{\alpha \beta }$ conservation equations \cite 
{wei}. In the ideal case we have :  
\begin{equation} 
\begin{array}{rcl} 
T_{\alpha \beta } & = & p\eta _{\alpha \beta }+(p+\varepsilon )U_\alpha 
U_\beta \\  
N_\alpha & = & nU_\alpha  
\end{array} 
\end{equation} 
where $\eta _{\alpha \beta }$ is the metric tensor, $p$ is the pressure,  
$\varepsilon $ is the energy density, n is the number of particles density  
and $U_\alpha :(\gamma \vec \beta ,\gamma )$ is the 4-velocity field. 
 
There are two ways of choosing the rest frame : in the Landau way, $U_\alpha  
$ is the energy transport velocity where $T_{i0}=0$ in the rest frame, while 
in the Eckart way $U_\alpha $ is the particle transport velocity where $%
N_i=0 $ in the rest frame. The dissipation contribution is introduced by 
redefining the energy-momentum and number of particle tensor by adding 
correction terms :  
\begin{equation} 
\begin{array}{rcl} 
T_{\alpha \beta } & = & p\eta _{\alpha \beta }+(p+\varepsilon )U_\alpha 
U_\beta +\Delta T_{\alpha \beta } \\  
N_\alpha & = & nU_\alpha +\Delta N_\alpha  
\end{array} 
\end{equation} 
In the Eckart frame $\Delta N_\alpha =0$, so the dissipation contribution is 
present only in the energy-momentum terms. In the following we choose the 
Eckart approach. The construction of the most general dissipation term  
$\Delta T_{\alpha \beta }$ is based on the positivity of the entropy 
production \cite{wei}:  
\begin{equation} 
\Delta T^{\alpha \beta }=-\eta H^{\alpha \gamma }H^{\beta \delta }W_{\gamma 
\delta }-\chi \left( H^{\alpha \gamma }U^\beta +H^{\beta \gamma }U^\alpha 
\right) Q_\gamma -\zeta H^{\alpha \beta }\partial _\gamma U^\gamma  
\end{equation} 
where we have shear tensor:  
\begin{equation} 
W_{\alpha \beta }=\partial _\beta U_\alpha +\partial _\alpha U_\beta -\frac 
23\eta _{\alpha \beta }\,\partial _\gamma U^\gamma  
\end{equation} 
heat-flow vector:  
\begin{equation} 
Q_\alpha =\partial _\alpha T+T\,\,U^\beta \,\partial _\beta \,U_\alpha  
\end{equation} 
T is the temperature and projection tensor on the hyperplane normal to  
$U_\alpha $  
\begin{equation} 
H_{\alpha \beta }=\eta _{\alpha \beta }+ U_\alpha \,U_\beta  
\end{equation} 
We identify $\chi ,\eta ,\zeta $ as the coefficients of heat conduction, 
shear viscosity and bulk viscosity. 
 
The conservation of the energy-momentum tensor $T_{\alpha \beta }$ gives us 
the main system of equations that controls the fluid dynamics:  
\begin{equation} 
\partial ^\alpha T_{\alpha \beta }=0  
\end{equation} 
 
A major obstacle in the application of symmetry analysis is the large number 
of tedious calculations usually involved. This is the reason to simplify the 
form of the equations in a reasonable way. Therefore, we are looking in the 
energy-momentum conservation equation for the powers of 4-velocity field $U$ 
because in the ultrarelativistic limit $U_\mu ^3\gg U_\mu ^2\gg U_\mu $;  
$\mu =1,2,3$ or $4$ (no summation), more precisely $U_\mu \sim \gamma 
\Rightarrow U_\mu ^2\sim \gamma ^2\Rightarrow U_\mu ^2\gg U_\mu \equiv 
\gamma ^2\gg \gamma $ and $\beta \simeq 1$. Taking only the higher power 
term and terms without $U_\mu $, we have \cite{phd}: 
\begin{equation} 
\partial ^\alpha T_{\alpha \beta }=\partial _\beta \left[ p+\left( \frac 
23\eta -\zeta \right) \partial ^\alpha U_\alpha \right] -2\chi \cdot 
\partial ^\alpha \left( U_\alpha U_\beta U^\gamma \partial _\gamma T\right)  
\end{equation} 
 
Performing calculations and in the end, taking only the highest contribution 
from the velocity field, we obtained from the energy-momentum conservation 
equation the following set of equations in 1+1 dimensions:  
\begin{equation} 
\label{eqs} 
\begin{array}{c} 
p_x+\left( \frac 23\eta -\zeta \right) \left( u_{xx}-v_{xt}\right) -2\chi 
\left( u^3T_{xx}+uv^2T_{tt}-2u^2vT_{xt}\right) =0 \\  
p_t+\left( \frac 23\eta -\zeta \right) \left( u_{xt}-v_{tt}\right) -2\chi 
\left( u^2vT_{xx}+v^3T_{tt}-2v^2uT_{xt}\right) =0  
\end{array} 
\end{equation} 
$U_\alpha :(\gamma \vec \beta ,\gamma )=(u,v)$ , $U^2=U_\alpha U^\alpha =-1$ 
, $u_{xx}=\partial ^2u/\partial x^2$, etc.. It is important to mention that 
from $\partial ^\alpha \left( U_\alpha U_\beta U^\gamma \partial _\gamma 
T\right) $ we took only the terms with $U_\mu ^3$, i.e. $\partial ^\alpha 
\left( U_\alpha U_\beta U^\gamma \partial _\gamma T\right) \rightarrow 
U_\alpha U_\beta U^\gamma \partial ^\alpha \partial _\gamma T$ and in this 
expression we neglected again terms containing velocity field, $u$ or $v$, 
at a power smaller than three. We also consider that the shear viscosity, 
the bulk viscosity and the heat conduction are constants; this is a major 
simplification for the symmetry group calculations. In fact they can be 
functions of temperature, for example in Weinberg's book \cite{wei}, where 
for a particular kind of fluid we have  
\begin{equation} 
\chi =\frac 43k_BT^3\tau \ ;\eta =\frac 4{15}k_BT^4\tau \ ;\zeta 
=4k_BT^4\tau \left[ \frac 13-\left( \frac{\partial p}{\partial \varepsilon }%
\right) \right] ^2  
\end{equation} 
where $\tau $ is the mean free time and $k_B$ is the Boltzmann constant. 
 
\section{Symmetry group of transformations and its Lie algebra} 
 
At the end of the last century, Lie considered the invariance of the 
differential equations under the transformation of dependent and independent 
variables \cite{lie}. Lie was able to classify and solve some types of 
ordinary as well as partial differential equations. In recent years, the 
symmetry methods have become more attractive, especially in the field of 
nonlinear dynamics \cite{olver,bluman}. 
 
The symmetry group of a system of differential equations is the largest 
local group of transformation acting on the independent and dependent 
variables of the system with the property that it transform solutions of the 
system to other solutions. We restrict our attention to local Lie group of 
symmetries, leaving aside problems involving discrete symmetries such as 
reflections. 
 
Let ${\cal S}$ be a system of differential equations. A symmetry-group of 
the system ${\cal S}$ is a local group of transformations ${\cal G}$ acting 
on an open subset ${\cal M}$ of the space of independent and dependent 
variables for the system with the property that whenever u=f(x) is a 
solution of ${\cal S}$, and whenever $g\cdot f$ is defined for $g\in {\cal G} 
$, then $u=g\cdot f(x)$ is also a solution of the system. 
 
The symmetry group infinitesimal generator is defined by :  
\begin{equation} 
\vec {{\cal V}}=\xi \partial _x+\tau \partial _t+\Phi \partial _u+\Psi 
\partial _v+\Gamma \partial _T+\Omega \partial _p  
\end{equation} 
and the first order prolongation of $\vec {{\cal V}}$ is:  
\begin{equation} 
\begin{array}{ll} 
pr^{(1)}\vec {{\cal V}}= & \xi \partial _x+\tau \partial _t+\Phi \partial 
_u+\Psi \partial _v+\Gamma \partial _T+\Omega \partial _p \\   
& +\Phi ^x\partial _{u_x}+\Phi ^t\partial _{u_t}+\Psi ^x\partial _{v_x}+\Psi 
^t\partial _{v_t} \\   
& +\Gamma ^x\partial _{T_x}+\Gamma ^t\partial _{T_t}+\Omega ^x\partial 
_{p_x}+\Omega ^t\partial _{p_t}  
\end{array} 
\end{equation} 
where, for example,  
\begin{equation} 
\Phi ^x=D_x(\Phi -\xi u_x-\tau u_t)+\xi u_{xx}+\tau u_{xt}  
\end{equation} 
and $D_x\Phi =\Phi _x+\Phi _uu_x+\Phi _vv_x+\Phi _TT_x+\Phi _pp_x$ is the 
total derivative. The second order prolongation of $\vec {{\cal V}}$ is 
defined by the following relation:  
\begin{equation} 
\begin{array}{rl} 
pr^{(2)}\vec {{\cal V}}= & \xi \partial _x+\tau \partial _t+\Phi \partial 
_u+\Psi \partial _v+\Gamma \partial _T+\Omega \partial _p \\   
& +\Phi ^x\partial _{u_x}+\Phi ^t\partial _{u_t}+\Psi ^x\partial _{v_x}+\Psi 
^t\partial _{v_t}+\Gamma ^x\partial _{T_x}+\Gamma ^t\partial _{T_t}+\Omega 
^x\partial _{p_x}+\Omega ^t\partial _{p_t} \\   
& +\Phi ^{xx}\partial _{u_{xx}}+\Phi ^{xt}\partial _{u_{xt}}+\Phi 
^{tt}\partial _{u_{tt}}+\Psi ^{xx}\partial _{v_{xx}}+\Psi ^{xt}\partial 
_{v_{xt}}+\Psi ^{tt}\partial _{v_{tt}} \\   
& +\Gamma ^{xx}\partial _{T_{xx}}+\Gamma ^{xt}\partial _{T_{xt}}+\Gamma 
^{tt}\partial _{T_{tt}}+\Omega ^{xx}\partial _{p_{xx}}+\Omega ^{xt}\partial 
_{p_{xt}}+\Omega ^{tt}\partial _{p_{tt}}  
\end{array} 
\end{equation} 
where, for example,  
\begin{equation} 
\Phi ^{xx}=D_x^2(\Phi -\xi u_x-\tau u_t)+\xi u_{xxx}+\tau u_{xxt}  
\end{equation} 
Suppose $\Delta _\nu (x,u^{(n)})=0,\nu =1,...,l,$ is a system of 
differential equations of maximal rank (meaning that the Jacobian matrix  
$J_{\bigtriangleup }(x,u^{(n)})=\left( \frac{\partial \bigtriangleup _\nu }{ 
\partial x^i},\frac{\partial \bigtriangleup _\nu }{\partial u_J^\alpha } 
\right) $ of with respect to all the variables $(x,u^{(n)})$ is of rank l 
whenever $\Delta (x,u^{(n)})=0$) defined over  
${\cal M}$$\subset $X$\times $U, where  
$u^{(n)}=\left( u,v,T,p,u_x,u_t,...,p_{tt}\right) $. If ${\cal G}$ is a local 
group of transformations acting on ${\cal M}$ and  
$pr^{(n)}\vec {{\cal V}}\left[\Delta _\nu (x,u^{(n)})\right] =0$,  
$\nu =1,...,l,$ whenever $\Delta (x,u^{(n)})=0$,  
for every infinitesimal generator ${\cal v}$ of ${\cal G}$, then  
${\cal G}$ is a symmetry group of the system. 
 
The standard procedure \footnote{ 
The method is well known and a good description can be found in \cite{olver}} 
is based on finding the infinitesimal coefficient functions $\xi ,\tau ,\Phi 
,\Psi ,\Gamma $ and $\Omega $. Applying $pr^{(2)}\vec {{\cal V}}$ on the 
system (\ref{eqs}) , we obtained the infinitesimal criterion  
\begin{equation} 
\begin{array}{c} 
\Omega ^x+\left( \frac 23\eta -\zeta \right) \left( \Phi ^{xx}-\Psi 
^{xt}\right) -2\chi \cdot \left( 3u^2\Phi \Gamma ^{xx}+2v\Phi \Psi \Gamma 
^{tt}-4u\Phi \Psi \Gamma ^{xt}\right) =0 \\  
\Omega ^x+\left( \frac 23\eta -\zeta \right) \left( \Phi ^{xx}-\Psi 
^{xt}\right) -2\chi \cdot \left( 2u\Phi \Psi \Gamma ^{xx}+3v^2\Psi \Gamma 
^{tt}-4v\Phi \Psi \Gamma ^{xt}\right) =0  
\end{array} 
\end{equation} 
Substituting the general formulae for $\Phi ^x,\Psi ^x,etc.$ and equating 
the coefficients of various monomials in the first and second order partial 
derivatives of $u,v,T$ and $p$, we find the defining equations. We wish to 
determine all possible coefficient functions $\xi ,\tau ,\Phi ,\Psi ,\Gamma $ 
and $\Omega $ by solving the defining equations system so that the 
corresponding one-parameter group $\exp (\varepsilon \vec {{\cal V}})$ is a 
symmetry group of the equations (\ref{eqs}). 
 
We will consider two cases: a) incompressible and b) compressible fluid.  
The basis of the corresponding Lie algebra is:  
\begin{equation} 
\begin{array}{ll} 
{\bf Incompressible} & {\bf Compressible} \\ V_1=\partial _x & V_1=\partial 
_x \\  
V_2=\partial _t & V_2=\partial _t \\  
V_3=\partial _T & V_3=\partial _T \\  
V_4=x\partial _T & V_4=x\partial _T \\  
V_5=t\partial _T & V_5=t\partial _T \\  
V_6=t\partial _x+x\partial _t-u\partial _u-v\partial _v & V_6=t\partial 
_x+x\partial _t-u\partial _u-v\partial _v \\  
V_7=x\partial _x+t\partial _t & V_7=x\partial _x+t\partial _t-p\partial _p 
\\  
V_8=u\partial _u+v\partial _v-2T\partial _T & V_8=u\partial _u+v\partial 
_v-2T\partial _T+p\partial _p \\   
& V_9=\partial _p  
\end{array} 
\end{equation} 
Using the following substitutions $x=\tau \cosh (\alpha )$ and $t=\tau \sinh 
(\alpha )$ we obtain that $t\partial _x+x\partial _t=\partial _\alpha $ - 
angle translation (rotation of the (x,t)-plane); if $u=\sinh (w)$ and  
$v=\cosh (w)\rightarrow w=\frac 12\log \frac{v+u}{v-u}$ ($w$ is the rapidity) 
we have $u\partial _v+v\partial _u=\partial _w$ which is a rapidity 
translation; it is important to mention that $V_6$ is a Lorentz 
transformation. 
 
\section{Solvable group and invariants} 
 
Because we have the Lie algebra of the system (\ref{eqs}) we want to know if 
the general solution of the system of differential equations can be found by 
quadratures. This thing is possible if the Lie group is solvable. The group  
${\cal G}$ is solvable if there exists a chain of Lie subgroups  
\begin{equation} 
{e}=G^{(0)}\subset G^{(1)}\subset \ldots \subset G^{(r-1)}\subset G^{(r)}=G  
\end{equation} 
such that for each k=1,$\dots $,r, ${\cal G}^{(k)}$ is a k-dimensional 
subgroup of ${\cal G}$ and ${\cal G}^{(k-1)}$ is a normal subgroup of  
${\cal G}^{(k)}$.  
A subgroup H is normal subgroup if $ghg^{-1}\in H$ whenever $g\in  
{\cal G}$ and $h\in H$. Equivalently, there is a chain of subalgebras  
\begin{equation} 
{e}=g^{(0)}\subset g^{(1)}\subset \ldots \subset g^{(r-1)}\subset g^{(r)}=g  
\end{equation} 
such that for k, $dimg^{(k)}=k$ and $g^{(k-1)}$ is a normal subalgebra of  
$g^{(k)}$:  
\begin{equation} 
[g^{(k-1)},g^{(k)}]\subset g^{(k-1)}  
\end{equation} 
 
The requirement for solvability is equivalent to the existence of a basis  
$\left\{ {V_1,\ldots ,V_r}\right\} $ of Lie algebra ${\it g}$ such that  
\begin{equation} 
[V_i,V_j]=\sum\limits_{k=1}^{j-1}c_{ij}^kV_k  
\end{equation} 
whenever $i<j$. 
\vskip -1cm 
\begin{table} 
\begin{center} 
\caption{Commutator table for the incompressible fluid algebra.}  
\footnotesize\rm 
\begin{tabular}{@{}lllllllll} \hline 
{\bf [\, , \,]} & V(1) & V(2) & V(3) & V(4) & V(5) & V(6) & V(7) & V(8) \\ \hline 
     V(1) & 0    & 0    & 0    & V(3) & 0    & V(2) & V(1) & 0    \\  
     V(2) & 0    & 0    & 0    &  0   & V(3) & V(1) & V(2) & 0    \\  
     V(3) & 0    & 0    & 0    &  0   & 0    & 0    & 0    &-2V(3) \\   
     V(4) &-V(3) & 0    & 0    &  0   & 0    &-V(5) &-V(4) &-2V(5) \\   
     V(5) & 0    &-V(3) & 0    &  0   & 0    &-V(4) &-V(5) &-2V(4) \\   
     V(6) &-V(2) &-V(1) & 0    & V(5) & V(4) & 0    & 0    & 0      \\   
     V(7) &-V(1) &-V(2) & 0    & V(4) & V(5) & 0    & 0    & 0      \\   
     V(8) & 0    & 0    &2V(3) &2V(4) &2V(5) & 0    & 0    & 0      \\ \hline  
\end{tabular} 
\end{center} 
\vskip -7mm 
\end{table} 
\vskip -1.5cm 
\begin{table} 
\begin{center} 
\caption{Commutator table for the compressible fluid algebra.} 
\footnotesize\rm 
\begin{tabular}{@{}llllllllll} \hline 
{\bf [\, , \,]} & V(1) & V(2) & V(3) & V(4) & V(5) & V(6) & V(7) & V(8) & V(9) \\ \hline 
     V(1) & 0    & 0    & 0    & V(3) & 0    & V(2) & V(1) & 0    & 0    \\  
     V(2) & 0    & 0    & 0    &  0   & V(3) & V(1) & V(2) & 0    & 0    \\  
     V(3) & 0    & 0    & 0    &  0   & 0    & 0    & 0    &-2V(3)& 0    \\  
     V(4) &-V(3) & 0    & 0    &  0   & 0    &-V(5) &-V(4) &-2V(5)& 0    \\  
     V(5) & 0    &-V(3) & 0    &  0   & 0    &-V(4) &-V(5) &-2V(4)& 0    \\  
     V(6) &-V(2) &-V(1) & 0    & V(5) & V(4) & 0    & 0    & 0    & 0    \\  
     V(7) &-V(1) &-V(2) & 0    & V(4) & V(5) & 0    & 0    & 0    & V(9) \\  
     V(8) & 0    & 0    &2V(3) &2V(4) &2V(5) & 0    & 0    & 0    &-V(9) \\  
     V(9) & 0    & 0    & 0    & 0    & 0    & 0    &-V(9) & V(9) & 0    \\ \hline 
\end{tabular} 
\end{center} 
\end{table} 
\vskip -1cm 
Looking at the commutator table of the Lie algebra we will see that the 
requirement of solvability is satisfy in both incompressible and 
compressible cases because we can construct the following chain of invariant 
sub-groups  
\begin{equation} 
\begin{array}{c} 
\{e\}=G^{[0]}\subset G^{[1]}\subset G^{[1,2]}\subset G^{[1,2,3]}\subset 
G^{[1,2,3,4]} \\  
\subset G^{[1,...,5]}\subset G^{[1,...,6]}\subset G^{[1,...,7]}\subset 
G^{[1,...,8]}=G  
\end{array} 
\end{equation} 
where $G^{[i,...,j]}$ is the subgroup generated by $V(i),...,V(j)$ for the 
incompressible fluid and  
\begin{equation} 
\begin{array}{c} 
\{e\}=G^{[0]}\subset G^{[1]}\subset G^{[1,2]}\subset G^{[1,2,3]}\subset 
G^{[1,...,4]}\subset G^{[1,...,5]} \\  
\subset G^{[1,...,6]}\subset G^{[1,...,7]}\subset G^{[1,...,8]}\subset 
G^{[1,...,9]}=G  
\end{array} 
\end{equation} 
for the compressible fluid.  
 
We use the method of characteristics to compute the invariants of the Lie 
algebra hopping that the reduced system, which can be obtained using the 
invariants of the group, will help us to solve the system of equations  
(\ref{eqs}). An n-th order differential invariant of a group G is a smooth 
function depending on the independent and dependent variables and their 
derivatives, invariant on the action of the corresponding n-th prolongation 
of G \cite{olver}. 
 
Suppose that we have the following generator:  
\begin{equation} 
V_i=\xi _i\partial _x+\tau _i\partial _t+\Phi _i\partial _u+\Psi _i\partial 
_v+\Gamma _i\partial _T+\Omega _i\partial _p  
\end{equation} 
A local invariant $\zeta $ of $V_i$ is a solution of the linear, homogeneous 
first order partial differential equation:  
\begin{equation} 
\label{invsol}V_i(\zeta )=\xi _i\partial _x\zeta +\tau _i\partial _t\zeta 
+\Phi _i\partial _u\zeta +\Psi _i\partial _v\zeta +\Gamma _i\partial _T\zeta 
+\Omega _i\partial _p\zeta =0  
\end{equation} 
The classical theory of such equations shows that the general solution of 
equation (\ref{invsol}) can be found by integrating the corresponding 
characteristic system of differential equations, which is  
\begin{equation} 
\frac{dx}{\xi _i}=\frac{dt}{\tau _i}=\frac{du}{\Phi _i}=\frac{dv}{\Psi _i}=%
\frac{dT}{\Gamma _i}=\frac{dp}{\Omega _i}  
\end{equation} 
Doing this integration we get, in this case, five invariants; we now 
re-express the next generator of Lie algebra in terms of these five 
invariants and then we perform another integration. We continue this 
calculation until we re-express and integrate the last generator; at this 
point we obtain a set of invariants that represent the system of independent 
invariants of this group. The system of invariants can be used to reduce the 
order of the original equations - constructing the reduced order system of 
equations. Doing this one can hope to find simple equations that can be 
integrated (for example \cite{olver}). 
 
Unfortunately our system of independent invariants is not so friendly and we 
can't simplify the form of the equations. We do not present here the 
invariants because of their unpleasant form and specially because they are 
useless in this particular application; the only important thing is that one 
of these invariants is $u^2-v^2$, which means that the unitarity of the 
velocity field is preserved. 
 
This method will be very well applied on the next section where the 
invariants are much more simple and we will use them to find the group 
invariant-solutions. 
In the next section we will focus on the incompressible fluid because the 
absence of the pressure term in our equations will allow us to obtain 
analytical solutions by integrating the equations; this is due to the number 
of dependent variable which decrease from four ($u,v,T$ and $p$) to three  
($u,v$ and $T$). 
 
\section{Group invariant-solutions} 
 
A solution of the system of partial differential equations is said to be  
${\cal G}$-invariant if it is unchanged by all the group transformations in  
${\cal G}$. In general, to each s-parameter subgroup ${\cal H}$ of the full 
symmetry group ${\cal G}$ of a system of differential equations, there will 
correspond a family of group-invariant solutions. Since there are almost 
always an infinite number of such subgroups, it is not usually feasible to 
list all possible group-invariant solutions to the system. We need an 
effective systematic means of classifying these solutions, leading to an 
optimal system of group-invariant solutions from which every other solution 
can be derived. Since elements ${\it g}$ $\in $ ${\cal G}$ not in the 
subgroup ${\cal H}$ will transform an ${\cal H}$-invariant solution to some 
other group-invariant solution, only those solutions not so related need to  
be listed in our optimal system. 
 
An optimal system of s-parameter subgroups is a list of conjugancy 
inequivalent s-parameter subgroups with the property that any other subgroup 
is conjugate to precisely one subgroup in the list (conjugacy map: h$%
\rightarrow $ghg$^{-1}$). 
 
Let ${\cal G}$ be a Lie group with Lie algebra ${\it g}$. For each $v\in  
{\it g}$, the adjoint vector $ad\ v$ at $w\in $ ${\it g}$ is  
\begin{equation} 
ad\ v|_w=[w,v]=-[v,w]  
\end{equation} 
Now we can reconstruct the adjoint representation $Ad\ {\cal G}$ of the Lie 
group by summing the Lie series  
\begin{equation} 
Ad(\exp (\varepsilon v))w=\sum\limits_{n=0}^\infty \frac{\varepsilon ^n}{n!}%
(ad\ v)^n(w)=w-\varepsilon [v,w]+\frac{\varepsilon ^2}2[v,[v,w]]-...  
\end{equation} 
obtaining the adjoint table.  
\vskip -0.7cm 
\begin{table} 
\caption{Adjoint table} 
\footnotesize\rm 
\begin{tabular}{@{}llll} \hline 
{\bf Ad} & {\bf V(1)} & {\bf V(2)} & {\bf V(3)}    \\ \hline 
{\bf V(1)} & V(1) & V(2) & V(3)   \\  
{\bf V(2)} & V(1) & V(2) & V(3)   \\  
{\bf V(3)} & V(1) & V(2) & V(3)  \\  
{\bf V(4)} & V(1)+$\varepsilon$V(3) & V(2) & V(3)  \\  
{\bf V(5)} & V(1) & V(2)+$\varepsilon$V(3) & V(3)  \\  
{\bf V(6)} & $e^{\varepsilon}$V(1) & $e^{\varepsilon}$V(2) & V(3) \\  
{\bf V(7)} & V(1) & V(2) & $e^{-2 \varepsilon}$V(3) \\  
{\bf V(8)} & $\cosh (\varepsilon)$ V(1) + $\sinh(\varepsilon)$ V(2)   
           & $\sinh (\varepsilon)$ V(1) + $\cosh(\varepsilon)$ V(2) & V(3) \\  
\end{tabular} 
\begin{tabular}{@{}llllll} \hline 
{\bf Ad}   & {\bf V(4)} & {\bf V(5)} & {\bf V(6)} & {\bf V(7)} & {\bf V(8)} \\ \hline 
{\bf V(1)} & V(4)-$\varepsilon$V(3) & V(5) & V(6)-$\varepsilon$V(2) & V(7)-$\varepsilon$V(1) & V(8) \\  
{\bf V(2)} & V(4) & V(5)-$\varepsilon$V(3) & V(6)-$\varepsilon$V(1) & V(7)-$\varepsilon$V(2) & V(8) \\  
{\bf V(3)} & V(4) & V(5) & V(6) & V(7) & V(8)+2$\varepsilon$V(3) \\  
{\bf V(4)} & V(4) & V(5) & V(6)+$\varepsilon$V(5) & V(7)+$\varepsilon$V(4) & 
     V(8)+2$\varepsilon$V(4) \\  
{\bf V(5)} & V(4) & V(5) & V(6)+$\varepsilon$V(4) & V(7)+$\varepsilon$V(5) & 
     V(8)+2$\varepsilon$V(5) \\  
{\bf V(6)} & $\cosh (\varepsilon)$V(4)-$\sinh (\varepsilon)$V(5) &  
     $\cosh (\varepsilon)$V(5)-$\sinh (\varepsilon)$V(4) &V(6)&V(7)&V(8) \\  
{\bf V(7)} & $e^{\varepsilon}$V(4)&$e^{\varepsilon}$V(5) &V(6)&V(7)&V(8) \\  
{\bf V(8)} & $e^{-2\varepsilon}$V(4)&$e^{-2\varepsilon}$V(5) &V(6)&V(7)&V(8) \\ \hline 
\end{tabular} 
\end{table} 
 
The optimal system of our equations (\ref{eqs}) is provided by those  
generated by  
\begin{equation} 
\begin{array}{rl} 
\fl 1) & V(6)+aV(7)+bV(8)=(ax+t)\partial _x+(at+x)\partial _t+(bu-v)\partial 
_u+(bv-u)\partial _v-2T\partial _T \\  
\fl 2) & aV(7)+V(8)=ax\partial _x+at\partial _t+u\partial _u+v\partial 
_v-2T\partial _T \\  
\fl 3) & aV(7)+V(6)=(ax+t)\partial _x+(at+x)\partial _t-v\partial _u-u\partial 
_v \\  
\fl 4) & aV(7)+V(6)\pm V(3)=(ax+t)\partial _x+(at+x)\partial _t-v\partial 
_u-u\partial _v\pm \partial _T \\  
\fl 5) & aV(8)+V(6)=t\partial _x+x\partial _t+(au-v)\partial _u+(av-u)\partial 
_v-2aT\partial _T \\  
\fl 6) & V(7)\pm V(3)=x\partial _x+t\partial _t\pm \partial _T \\  
\fl 7) & V(8)\pm V(1)=\pm \partial _x+u\partial _u+v\partial _v-2T\partial _T\ 
;V(8)\pm V(2)=\pm \partial _t+u\partial _u+v\partial _v-2T\partial _T \\  
\fl 8) & V(6)\pm V(3)=t\partial _x+x\partial _t-v\partial _u-u\partial _v\pm 
\partial _T \\  
\fl 9) & V(1)+V(4)+aV(5)=\partial _x+(x+at)\partial _T\ 
;V(1)+V(5)+aV(4)=\partial _x+(t+ax)\partial _T\  \\  
\fl 10) & V(2)+V(4)+aV(5)=\partial _t+(x+at)\partial _T\ 
;V(2)+V(5)+aV(4)=\partial _t+(t+ax)\partial _T\  \\  
\fl 11) & V(4)+V(1)+aV(2)=\partial _x+a\partial _t+x\partial _T\ 
;V(4)+V(2)+aV(1)=a\partial _x+\partial _t+x\partial _T \\  
\fl 12) & V(5)+V(1)+aV(2)=\partial _x+a\partial _t+t\partial _T\ 
;V(5)+V(2)+aV(1)=a\partial _x+\partial _t+t\partial _T \\  
\fl 13) & V(1)+V(4)=\partial _x+x\partial _{T\ };V(1)+V(5)=\partial _x+t\partial 
_{T\ } \\  
\fl 14) & V(2)+V(4)=\partial _t+x\partial _{T\ };V(2)+V(5)=\partial _t+t\partial 
_{T\ } \\  
\fl 15) & V(i),i=1,...,8 
\end{array} 
\end{equation} 
where a and b are arbitrary constants. 
 
In this case, of incompressible fluid, the equations (\ref{eqs}) are  
\begin{equation} 
\begin{array}{c} 
u_{xx}-v_{xt}-k\cdot \left( u^3T_{xx}+uv^2T_{tt}-2u^2vT_{xt}\right) =0 \\  
u_{xt}-v_{tt}-k\cdot \left( u^2vT_{xx}+v^3T_{tt}-2uv^2T_{xt}\right) =0  
\end{array} 
\end{equation} 
where $k=2\chi /\left( \frac 23\eta -\zeta \right) $; we can re-write them 
in the following form:  
\begin{equation} 
\label{lasteqs} 
\begin{array}{c} 
v\left( u_{xx}-v_{xt}\right) -u(u_{xt}-v_{tt})=0 \\  
u_{xx}-v_{xt}-k\cdot \left( u^3T_{xx}+uv^2T_{tt}-2u^2vT_{xt}\right) =0  
\end{array} 
\end{equation} 
 
Finally, we will concentrate our attention on the classification of the 
group-invariant solutions, but we will focus on the equations that can be 
solved analytically. 
 
1) In terms of the invariants $w=\rho +\frac 1a\log (ax+t),\beta 
=T(at+x)^{-2b/a}$ and $y=(t+x)^{a+1}(t-x)^{1-a}$ , where $\rho =0.5\log  
\frac{v+u}{v-u}$ the reduced system of equations are very complicated but  
if we choose $a=\pm 1$ they are  
\begin{equation} 
\begin{array}{c} 
w_{yy}\pm w_y^2+1.5y^{-1}w_y=0 \\  
2y^2\beta _{yy}\pm \left( 4b\pm 1\right) y\beta _y+b\left( 2b\mp 1\right) 
\beta =0  
\end{array} 
\end{equation} 
and the solutions are \cite{kamke}  
\begin{equation} 
T=\cases{  
c_1+c_2\left( t\pm x\right) \\  
c_1+2c_2\log \left( t\pm x\right) \\  
\left( t\pm x\right) \left[ c_1+2c_2\log \left( t\pm x\right) \right] \\} 
\end{equation} 
\begin{equation}  
\mid {\it \vec v}\mid =\mp \tanh \left\{ \log \left[ \pm 2c_3\mp c_4\left( 
t\pm x\right) \right] \right\}  
\end{equation} 
where $c_1,...,c_4$ are constants. 
 
2) using $\rho =0.5\log \frac{v+u}{v-u}$ we obtained in terms of the 
invariant $y=x/t$ the following reduced equation  
\begin{equation} 
\rho _{yy}+\rho _y^2\frac{y+\tanh (\rho) }{1+y\tanh (\rho) } 
+\rho _y\frac{2\tanh (\rho) }{1+y\tanh (\rho) }=0  
\end{equation} 
which is an equation that has to solved using numerical codes. The second 
reduced equation is much more complicated but it can be solved once we have 
the solution from the first equation. 
 
3) the invariants are $w=\rho +\frac 1a\log (ax+t)$ and  
$y=(t+x)^{a+1}(t-x)^{1-a}$ and for the same reason mentioned above we 
considered a=$\pm 1$ obtaining  
\begin{equation} 
\begin{array}{c} 
w_{yy}\pm w_y^2-0.5y^{-1}w_y\mp 0.5y^{-2}=0 \\  
2yT_{yy}+T_y=0  
\end{array} 
\end{equation} 
The temperature solution is $T=c_1+2c_2\sqrt{y}$;  
for the second equation we need numerical codes. 
 
4) the invariants are $w=\rho +\frac 1a\log (ax+t)$ ,  
$y=(t+x)^{a+1}(t-x)^{1-a}$ and $\beta =\pm T-\frac 1a\log (at+x)$; for  
a=$\pm 1$ we have  
\begin{equation} 
\begin{array}{c} 
w_{yy}\mp w_y^2+1.5y^{-1}w_y=0 \\  
\beta _{yy}+0.5y^{-1}\beta _y\mp 0.25y^{-2}=0  
\end{array} 
\end{equation} 
and the solutions are  
\begin{equation} 
\begin{array}{c} 
T=-\frac{5/4}{\left( t\pm x\right) ^4}+\frac{0.5}{\left( t\pm x\right) ^2}\pm 
2c_1\left( t\pm x\right) \pm c_2\pm \log \left( t\pm x\right) \\  
\mid {\it \vec v}\mid =\mp \tanh \{\log [\pm 2c_1\mp c_2(t\pm x)]\}  
\end{array} 
\end{equation} 
 
5) the reduced system of equations is very unpleasant and we need numerical 
codes 
 
6) here we have the same problem as in the case number 2) (the reduced 
system of equations is the same as in the second case) 
 
7), 8), 9) and 10) because of the form of the invariants we can not construct 
the reduced system of equations 
 
11) the invariants and the solutions are $\beta =T-\frac{xt}a$ and $y=ax-t$ 
and respectively, for a=$\pm 1$:  
\begin{equation} 
\begin{array}{c} 
T=\pm xt+0.25\left( \mp x-t\right) ^2- 
\frac{1/12}{c_1^2\left[ c_1\left( \mp x-t\right) +c_2\right] ^2}+c_3\left( 
\mp x-t\right) +c_4 \\ \mid {\it \vec v}\mid =\pm \tanh \left\{ \log \left[ 
c_1+c_2\left( x\mp t\right) \right] \right\}  
\end{array} 
\end{equation} 
For the second transformation we have the following invariants $\beta =T-xt$,  
$y=at-x$ and solutions, for a=$\pm 1$:  
\begin{equation} 
\begin{array}{c} 
T=xt+0.25\left( \mp t-x\right) ^2- 
\frac{1/12}{c_1^2\left[ c_1\left( \mp t-x\right) +c_2\right] ^2}+c_3\left( 
\mp t-x\right) +c_4 \\ \mid {\it \vec v}\mid =\pm \tanh \left\{ \log \left[ 
\mp c_1+c_2\left( t\mp x\right) \right] \right\}  
\end{array} 
\end{equation} 
 
12) the invariants are $y=ax-t$ , $\beta =T-xt$ and the solutions are, for  
a=$\pm 1$:  
\begin{equation} 
\begin{array}{c} 
T=xt+0.25\left( \mp x-t\right) ^2- 
\frac{1/12}{c_1^2\left[ c_1\left( \mp x-t\right) +c_2\right] ^2}+c_3\left( 
\mp x-t\right) +c_4 \\ \mid {\it \vec v}\mid =\pm \tanh \left\{ \log \left[ 
c_1+c_2\left( x\mp t\right) \right] \right\}  
\end{array} 
\end{equation} 
For the second transformation we have the following invariants   
$\beta =T-\frac{xt}a$, $y=at-x$ and solutions, for a=$\pm 1$: 
\begin{equation} 
\begin{array}{c} 
T=\pm xt+0.25\left( \mp t-x\right) ^2- 
\frac{1/12}{c_1^2\left[ c_1\left( \mp t-x\right) +c_2\right] ^2}+c_3\left( 
\mp t-x\right) +c_4 \\ \mid {\it \vec v}\mid =\pm \tanh \left\{ \log \left[ 
\mp c_1+c_2\left( t\mp x\right) \right] \right\}  
\end{array} 
\end{equation} 
 
13) and 14) because of the form of the invariants we can't construct the 
reduced system of equations 
 
15) consider the transformation $V(6)=x\partial_x+t\partial_t$, 
the reduced equations are the same as in the case 2); in all the other  
transformations we can not obtain reduced equations. 
 
\section{Summary and conclusions} 
The results of the symmetry group analysis of the  
energy-momentum tensor conservation equation for the imperfect fluid flow 
can be summarized by the following remarks: 
\begin{itemize} 
\item The ultrarelativistic limit was implemented in a simple analytical  
manageable way on the equations of motion. 
\item The local Lie symmetries of the equations were presented. 
\item The optimal system of transformation was calculated. 
\item We present all the analytical solutions of the reduced system of  
equations. 
\item The equation that has to be solved numerically was written in the  
reduced form using the invariants of the transformation. 
\item These analytical solutions can be very useful for the investigation of 
different physical systems where the dissipative processes are important. 
One of them is the relativistic heavy ion collisions where this kind 
of relativistic hydrodynamic equations are usually applied \cite{bjo}. 
\end{itemize} 
 
There are some questions that have not been addressed in this paper: 
\begin{quote} 
- we have not take into account the pressure and the energy density \par 
- there are also other terms with smaller power of the velocity field that 
were neglected \par 
- only longitudinal expansion was consider and the three-dimensional radial 
expansion of the fluid have not been discussed \par 
\end{quote} 
 
We will give short answers to the questions mention above:  
\begin{quote} 
- for the first problem we need a relation between the pressure and the  
energy density which can be used for dissipative systems \par 
- the second one will be the goal of our future analyses \par 
- the last one needs numerical codes and a particular physical system with  
known initial conditions \par 
\end{quote} 
 
We demonstrated the application of the Lie symmetry method on some 
particular equations proving that the differential invariants can help us to 
simplify very much the task of finding the solutions of some given 
differential equations. 
 
The Lie group approach in its general form is particularly effective since 
it furnishes both general Lie symmetries and all their invariants in a  
constructive way. 
 
We find that the application of this method will give us a straightforward 
way to decide the question of integrability. 
It appears that cases of exact solutions of differential equations 
are based on the use of symmetry of these equations with respect to certain 
transformations. 
 
\ack 
We wish to thank L. Anton for fruitful discussions and suggestions. We also 
thank Prof. M. Visinescu for his constant support and help.

\end{document}